\documentclass{appolb}
\usepackage{epsfig}
\def\prt{\partial}

\begin{document}
\title{Chiral symmetry, strangeness and nuclear matter
\thanks{Invited talk presented at the Mazurian Lake Conference 2003}%
}
\author{Matthias F.M. Lutz
\address{Gesellschaft f\"ur Schwerionenforschung GSI,
Postfach 110552\\ D-64220 Darmstadt, Germany}
}
\maketitle
\begin{abstract}
In this talk we review the important role played by chiral SU(3) symmetry
in hadron physics. Exciting new results on the formation of baryon resonances
as implied by chiral coupled-channel dynamics are presented and discussed.
The results are the consequence of progress made in formulating a
consistent effective field theory for the meson-baryon scattering processes
in the resonance region. Strangeness channels are found to play a decisive role
in the formation of resonances. As a further application of chiral coupled-channel
dynamics the properties of antikaons and hyperon resonances in cold nuclear matter
are reviewed.
\end{abstract}


\section{Introduction}

The meson-baryon scattering processes are an important test for effective field theories
which aim at reproducing QCD at small energies, where the effective degrees of freedom are
hadrons rather than quarks and gluons. In this talk we focus on the strangeness sectors, because
here the acceptable effective field theories were until recently much less developed and also the
empirical data set still leaves much room for different theoretical interpretations. In the near
future the new DA$\Phi$NE facility at Frascati could deliver new data on kaon-nucleon scattering
\cite{DAPHNE} and therewith help to establish a more profound understanding of the role played
by the $SU(3)$ flavor symmetry in hadron interactions. A reliable construction
of the meson-baryon scattering amplitudes is of major importance since they play a central
role in the study of meson properties in cold nuclear matter.

The task to construct a systematic effective field theory for the meson-baryon scattering
processes in the resonance region is closely linked to the fundamental question
as to what is the 'nature' of baryon resonances. Here we support the conjecture
\cite{LK01,LK02,LWF02,LH02} that baryon resonances not belonging to the large-$N_c$ ground
states are generated dynamically by coupled-channel dynamics
\cite{Wyld,Dalitz,Ball,Rajasekaran,Wyld2,sw88}. For a comprehensive
discussion of this issue we refer to \cite{LH02}. This conjecture was the basis of
the phenomenological model \cite{LWF02}, which generated successfully non-strange
s- and d-wave resonances by coupled-channel dynamics describing a large body of pion and
photon scattering data. In recent works \cite{LK01,LK02,LK00,Granada,KL03}, which will be reviewed
here, it was shown that chiral dynamics as implemented by the $\chi-$BS(3) approach
\cite{LK01,LK02,LH02} provides a parameter-free prediction for the
existence of a wealth of strange and non-strange s- and d-wave wave baryon resonances.

In the second part of this talk the application of chiral coupled-channel dynamics
to nuclear matter properties of the antikaon, the $\Lambda (1405)$ s-wave resonance and
the $\Sigma(1385)$ p-wave resonance will we discussed. An attractive modification of the
antikaon spectral function was already anticipated in the 70's by the many K-matrix analyses
of antikaon-nucleon scattering (see e.g. \cite{A.D.Martin}) which predicted considerable
attraction in the subthreshold s-wave
$K^-$ nucleon scattering amplitudes. In conjunction with the low-density theorem \cite{njl-lutz}
this leads to an attractive antikaon spectral
function in nuclear matter. As was pointed out first in \cite{ml-sp} the realistic evaluation of
the antikaon self energy in nuclear matter requires a self consistent scheme. The
feedback effect of an attractive antikaon spectral function on the antikaon-nucleon scattering
process was found to be important for the hyperon resonance structure in nuclear matter.
We present and discuss up-to-date results based on the chiral-coupled channel analysis of meson-baryon
scattering data that included s-, p- and d-wave contributions \cite{LK02,LuKor}.

\section{Effective field theory of chiral coupled-channel dynamics}

The starting point to describe the meson-baryon scattering process
is the chiral SU(3) Lagrangian (see e.g.\cite{Krause,LK02}). A systematic approximation
scheme arises due to a successful scale separation justifying the chiral
power counting rules \cite{book:Weinberg}. Our effective field theory of the meson-baryon
scattering processes is based on the assumption that the scattering amplitudes are
perturbative at subthreshold energies with the expansion parameter
$Q/ \Lambda_{\chi}$. The small scale $Q$ is to be identified with any small
momentum of the system. The chiral symmetry breaking scale is
$$\Lambda_\chi \simeq 4\pi f \simeq 1.13 \;{\mbox GeV}\,, $$ with the parameter $f\simeq 90$ MeV
determined by the pion decay process. Once the available
energy is sufficiently high to permit elastic two-body scattering a further typical
dimensionless parameter $m_K^2/(8\,\pi f^2) \sim 1$ arises \cite{LK00,LK01,LK02}. Since this
ratio is uniquely linked to two-particle reducible diagrams it is sufficient to sum
those diagrams keeping the perturbative expansion of all irreducible
diagrams, i.e. the  coupled-channel Bethe-Salpeter equation has to be solved. This is the
basis of the $\chi$-BS(3) approach developed in \cite{LK00,LK01,LK02}.

At leading order in the chiral expansion one encounters the famous
Weinberg-Tomozawa \cite{Wein-Tomo} interaction,
\begin{eqnarray}
\mathcal{L}_{WT}&=&
 \frac{i}{8\, f^2}\, {\rm tr}\, \Big((\bar B
\,\gamma^\mu\,\, B) \cdot
 [\Phi,(\prt_\mu \Phi)]_-  \Big)
\nonumber\\
&+&\frac{3\,i}{8\, f^2}\, {\rm tr}\, \Big((\bar B_\nu
\,\gamma^\mu\,\, B^\nu) \cdot
 [\Phi,(\prt_\mu \Phi)]_-  \Big)
 \,,
 \label{WT-term}
\end{eqnarray}
where we dropped additional structures that do not contribute to the on-shell scattering process
at tree level. The terms in (\ref{WT-term}) constitute the leading order s-wave interaction
of Goldstone bosons ($\Phi $) with the baryon-octet ($B$) and baryon-decuplet ($B_\mu$) states.
The octet and decuplet fields, $\Phi, B$ and $B_\mu$, posses an appropriate matrix structure
according to their SU(3) tensor representation.

The scattering process is described by the amplitudes that follow as solutions of the
Bethe-Salpeter equation,
\begin{eqnarray}
T(\bar k ,k ;w ) &=& V(\bar k ,k ;w )
+\int\!\! \frac{d^4l}{(2\pi)4}\,V(\bar k ,
l;w )\, G(l;w)\,T(l,k;w )\;,
\nonumber\\
G(l;w)&=&-i\,D(\half\,w-l)\,S( \half\,w+l)\,,
\label{BS-coupled}
\end{eqnarray}
where we suppress the coupled-channel structure for simplicity. The meson and
baryon propagators,  $D(q)$ and $S(p)$, are used in the notation of \cite{LWF02}.
We apply the convenient kinematics:
\begin{eqnarray}
w = p+q = \bar p+\bar q\,,
\quad k= \half\,(p-q)\,,\quad
\bar k =\half\,(\bar p-\bar q)\,,
\label{def-moment}
\end{eqnarray}
where $q,\,p,\, \bar q, \,\bar p$ the initial and final meson and baryon 4-momenta.
The Bethe-Salpeter scattering equation is recalled for the case of meson baryon-octet
scattering. An analogous equation holds for meson baryon-decuplet scattering process
(see e.g. \cite{LWF02}). Referring to the detailed discussion given in \cite{LK02} we assume a
systematic on-shell reduction of the Bethe-Salpeter interaction kernel
leading to the effective interaction $V$ used in (\ref{BS-coupled}).
The latter is is expanded according to chiral power counting rules.
The scattering amplitude $T(\bar k,k;w)$ decouples into various sectors
characterized by isospin ($I$) and strangeness ($S$) quantum numbers. In the
case of meson baryon-octet and baryon-decuplet scattering the following channels are relevant
\begin{eqnarray}
&& (I,S)_{[8 \otimes 8]} =  (0,-3), (1,-3), (\frac{1}{2},-2), (\frac{3}{2},-2) ,
(0,-1),  \nonumber\\
&& \qquad \qquad (1,-1),
(2,-1), (\frac{1}{2},0),(\frac{3}{2},0), (0,1),(1,1) \,,
\nonumber\\
&& (I,S)_{[8 \otimes 10]} = (\frac{1}{2},-4), (0,-3), (1,-3), (\frac{1}{2},-2), (\frac{3}{2},-2) ,
(0,-1), \nonumber\\
&& \qquad \qquad  (1,-1),
(2,-1),(\frac{1}{2},0),(\frac{3}{2},0), (\frac{5}{2},0), (1,1),(2,1) \,.
\label{sectors-10}
\end{eqnarray}
Following the $\chi$-BS(3) approach developed in \cite{LK02,LWF02} the effective
interaction kernel is decomposed into a set of covariant projectors that have well
defined total angular momentum, $J$, and parity, $P$,
\begin{eqnarray}
&& V(\bar k ,k ;w )  = \sum_{J,P}\,V^{(J,P)}(\sqrt{s}\,)\,
{\mathcal Y}^{(J,P)}(\bar q, q,w) \,.
\label{def-proj}
\end{eqnarray}
The merit of the projectors is that they decouple the Bethe-Salpeter
equation (\ref{BS-coupled}) into orthogonal sectors labelled by the total
angular momentum, $J$, and parity, $P$. We insist on the renormalization condition,
\begin{eqnarray}
T^{(I,S)}(\bar k,k;w)\Big|_{\sqrt{s}= \mu (I,S)} =
V^{(I,S)}(\bar k,k;w)\Big|_{\sqrt{s}= \mu (I,S)} \,,
\label{ren-cond}
\end{eqnarray}
together with the natural choice for the subtraction points,
\begin{eqnarray}
&& \mu(I,+1)=\mu(I,-3)={\textstyle{1\over 2}}\,(m_\Lambda+ m_\Sigma) \,,
\quad \mu(I,0)=m_N\,, \quad
\nonumber\\
&&  \mu(0,-1)=m_\Lambda,\quad \mu(1,-1)=m_\Sigma\,, \quad
\mu(I,-2)= \mu(I,-4)= m_\Xi
 \label{eq:sub-choice}
\end{eqnarray}
as explained in detail in \cite{LK02}. The renormalization condition reflects
the basic assumption our effective field theory is based on, namely that
at subthreshold energies the scattering amplitudes can be
evaluated in standard chiral perturbation theory. This is achieved
by supplementing (\ref{BS-coupled}) with (\ref{ren-cond},\ref{eq:sub-choice}).  The
subtraction points
(\ref{eq:sub-choice}) are the unique choices that protect the s-channel
baryon-octet masses manifestly in the p-wave $J={\textstyle{1\over 2}}$
scattering amplitudes. The merit of the scheme \cite{LK00,LK01,LK02} lies in the
property that for instance the $K \,\Xi$ and $\bar K\,\Xi $
scattering amplitudes match at $\sqrt{s} \sim m_\Xi $
approximately as expected from crossing symmetry. In \cite{LK02} we suggested
to glue s- and u-channel unitarized scattering amplitudes at subthreshold energies.
This construction reflects our basic assumption that diagrams showing an s-channel
or u-channel unitarity cut need to be summed to all orders at least at energies close to
where the diagrams develop their imaginary part. By construction, a glued scattering amplitude
satisfies crossing symmetry exactly at energies where the scattering process takes
place. At subthreshold energies crossing symmetry is implemented
approximatively only, however, to higher and higher accuracy when more chiral correction
terms are considered. Insisting on the renormalization condition
(\ref{ren-cond},\ref{eq:sub-choice}) guarantees that subthreshold amplitudes match smoothly
and therefore the final 'glued' amplitudes comply with the crossing-symmetry constraint
to high accuracy. The natural subtraction points
(\ref{eq:sub-choice}) can also be derived if one incorporates photon-baryon
inelastic channels. Then additional constraints arise. For instance the
reaction $\gamma \,\Xi \to \gamma \,\Xi $,
which is subject to a crossing symmetry constraint at threshold, may
go via the intermediate states $\bar K \,\Lambda $ or $\bar K \,\Sigma $.

The perturbative nature of subthreshold amplitudes, a crucial
assumption of the $\chi$-BS(3) approach proposed in \cite{LK00,LK01,LK02}, is
not necessarily true in phenomenological coupled-channel schemes
in \cite{ksw95,grnpi,grkl,Oset-prl,Oset-plb,Jido03}. Using the subtraction scales
as free parameters, as advocated in \cite{Oset-prl,Oset-plb,Jido03}, may be viewed as
promoting the counter terms of chiral order $Q^3$ to be unnaturally large. If the
subtraction scales are chosen far away from their natural values (\ref{eq:sub-choice})
the resulting loop functions are in conflict with chiral power counting rules \cite{LK00}.
Though unnaturally large $Q^3$ counter terms can not be excluded from first principals
one should check such an assumption by studying corrections terms systematically. A detailed
test of the naturalness of the $Q^3$ counter terms was performed within the $\chi$-BS(3)
scheme \cite{LK02} demonstrating good convergence in the channels studied
without any need for promoting the counter terms of order $Q^3$. Possible
correction terms in the approach followed in \cite{Oset-prl,Oset-plb,Jido03}
have so far not been studied systematically for meson-baryon scattering. Moreover,
if the scheme advocated in \cite{Oset-prl,Oset-plb,Jido03} were applied in all eleven isospin
strangeness sectors with $J^P=\frac{1}{2}^-$  a total number of 26
subtraction parameters arise. This should be compared with the only ten counter terms of chiral
order $Q^3$ contributing to the on-shell scattering amplitude at that order \cite{LK02}.
Selecting only the operators that are leading in the large-$N_c$ limit of QCD out of the
ten $Q^3$ operators only four survive \cite{LK02}. We conclude that it would be
inconsistent to apply the approach used in \cite{Oset-prl,Oset-plb,Jido03} in all isospin
strangeness channels without addressing the above mismatch of parameters. Our scheme has
the advantage over the one in~\cite{Oset-prl,Oset-plb,Jido03} that once the parameters
describing subleading effects are determined in a subset of sectors one has immediate
predictions for all sectors $(I,S)$. A mismatch of the number of parameters is avoided
altogether since the $Q^3$ counter terms enter the effective interaction kernel directly.

Given the subtraction scales (\ref{eq:sub-choice}) the leading order calculation is
parameter free. Of course chiral correction terms do lead to further so far unknown
parameters  which need to be adjusted to data. Within the $\chi-$BS(3) approach such
correction terms enter the effective interaction kernel $V$ rather than leading to
subtraction scales different from (\ref{eq:sub-choice}) as it is assumed
in~\cite{Oset-prl,Oset-plb,Jido03}. In particular the leading
correction effects are determined by the counter terms of chiral
order $Q^2$. The effect of altering the subtraction scales away
from their optimal values (\ref{eq:sub-choice}) can be compensated
for by incorporating counter terms in the chiral Lagrangian that
carry order $Q^3$.

\section{Baryon resonances from chiral SU(3) symmetry}

There is a long standing controversy to what is the nature of s-wave baryon resonances.
Before the event of the quark model several such states have been successfully generated
in terms of coupled-channels dynamics \cite{Wyld,Dalitz,Ball,Rajasekaran,Wyld2}. These
early calculations are closely related to modern approaches based on the leading-order
chiral SU(3) Lagrangian. The interaction used in \cite{Wyld,Dalitz,Ball,Rajasekaran,Wyld2}
matches the Weinberg-Tomozawa interaction (\ref{WT-term}) if expanded in a Taylor
series \cite{sw88}. The main difference of the early attempts from computations based on
the chiral Lagrangian is the way the coupled-channel scattering equation is regularized
and renormalized. The
crucial advance over the last years in this field is therefore a significant improvement
of the systematics, i.e. how to implement corrections terms into coupled-channel dynamics.
In the SU(6) quark-model approach s-wave resonances belong to a $70$-plet, that
contains many more resonance states \cite{Schat}. An interesting question arises: what is
the role played by the d-wave resonances belonging to the very same $70$-plet as the s-wave
resonances. Naively one may expect that chiral dynamics does not make firm
predictions for d-wave resonances since the meson baryon-octet
interaction in the relevant channels probes a set of
counter terms presently unknown. However, this is not
necessarily so. Since a d-wave baryon resonance couples to s-wave
meson baryon-decuplet states chiral symmetry is quite predictive
for such resonances under the assumption that the
latter channels are dominant. This is in full analogy to the
analysis of the s-wave resonances  \cite{Wyld,Dalitz,Ball,Rajasekaran,Wyld2,LK00,LK02,Granada,Jido03}
that neglects the effect of the contribution of d-wave meson baryon-decuplet
states. The empirical observation that the d-wave resonances
$N(1520)$, $N(1700)$ and $\Delta (1700)$ have large branching
fractions ($> 50 \% $) into the inelastic $N \pi \pi$ channel, even
though the elastic $\pi N$ channel is favored by phase space,
supports our assumption.

We begin with a discussion of the s- and d-wave baryon resonance spectrum that arises in
the SU(3) limit. The latter is not defined uniquely depending on the magnitude of the current
quark masses, $m_u=m_d=m_s$. We consider two scenarios \cite{Granada,KL03}. In the 'light' SU(3)
limit the current quark masses are chosen such that one obtains $m_\pi= m_K=m_\eta =140$ MeV.
The second case, the 'heavy' SU(3) limit, is characterized by $m_\pi= m_K=m_\eta =500$ MeV.
In the SU(3) limit meson baryon-octet scattering is classified according to,
\begin{eqnarray}
&& 8 \otimes 8 = 27 \oplus \overline{10} \oplus 10 \oplus 8 \oplus 8 \oplus 1 \,.
\label{8-8-decom}
\end{eqnarray}
The leading order chiral Lagrangian predicts attraction in the
two octet and the singlet channel but repulsion in the 27-plet and decuplet channels.
As a consequence in the 'heavy' SU(3) limit the chiral dynamics predicts two degenerate
octet bound states together with a non-degenerate singlet state
\cite{Wyld,Dalitz,Ball,Rajasekaran,Wyld2,Granada,Jido03,Granada,KL03}.
In the 'light' SU(3) limit all states disappear leaving no clear signal in any of the
speed plots.
In the $J^P=\frac{3}{2}^-$ sector the Weinberg-Tomozawa interaction is attractive in
the octet, decuplet and 27-plet channel, but repulsive in the 35-plet channel,
\begin{eqnarray}
&&8 \otimes 10 = 35 \oplus 27 \oplus 10 \oplus 8\,.
\label{8-10-decom}
\end{eqnarray}
Therefore one may expect resonances or bound states in the former channels.
Indeed, in the 'heavy' SU(3) limit we find $72=4\times (8+10)$ bound states in this sector
forming an octet and decuplet representation of the SU(3) group. We do not find a
27-plet-bound state reflecting the weaker attraction in that channel. However, if we
artificially increase the amount of attraction by about 40 $\%$ by lowering the value of $f$
in the Weinberg-Tomozawa term, a clear bound state arises in this channel also. A contrasted
result is obtained if we lower the meson masses down to the pion mass arriving at
the 'light' SU(3) limit. Then we find neither bound nor resonance octet or decuplet states.
This pattern is a clear prediction of chiral couple-channel dynamics
which should be tested with unquenched QCD lattice simulations \cite{dgr}.

\begin{figure}[t]
\begin{center}
\includegraphics[width=12.0cm,clip=true]{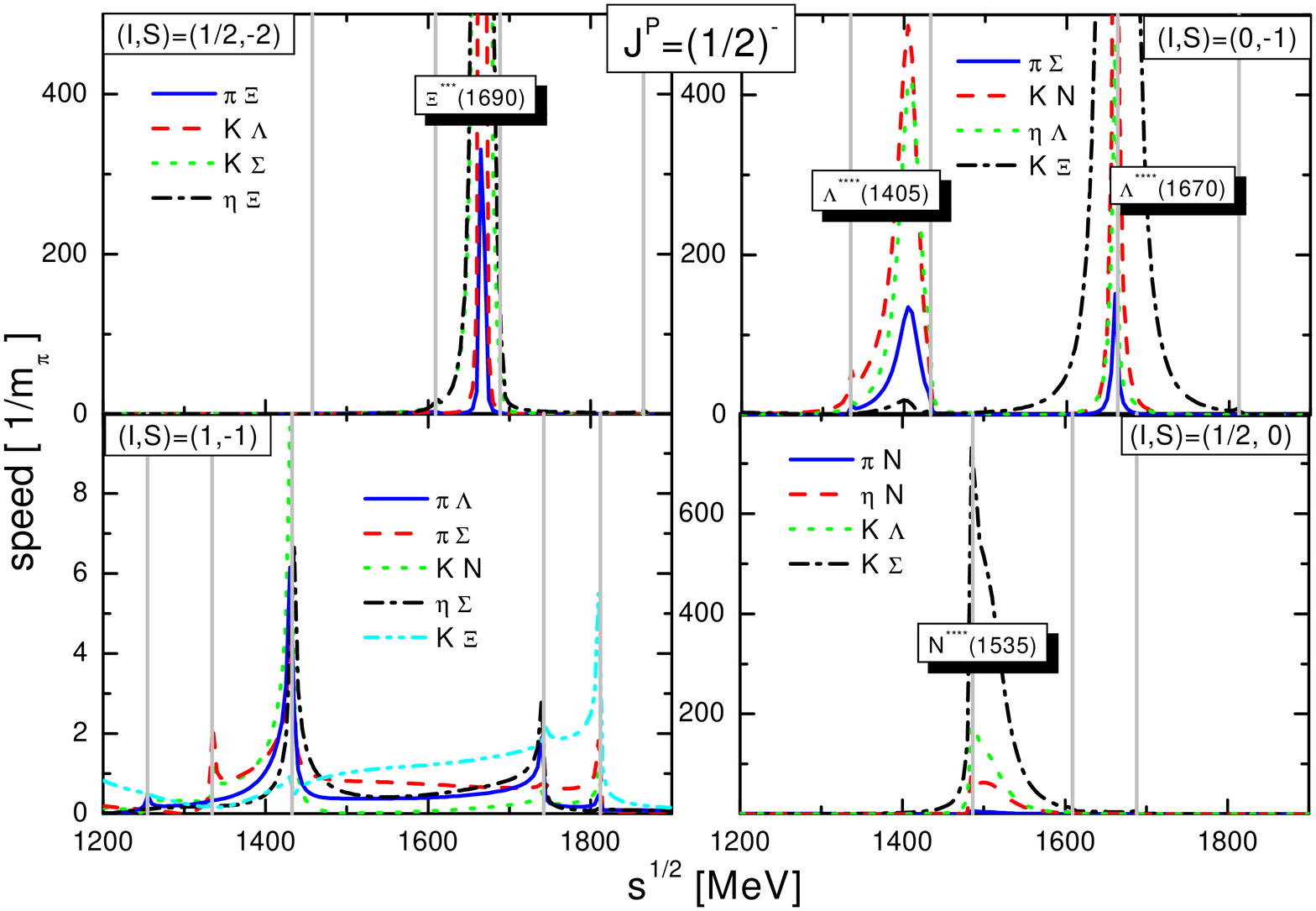}
\end{center}
\caption{Diagonal speed plots of the $J^P=\frac{1}{2}^-$ sector. The vertical lines show the
opening of inelastic meson baryon-decuplet channels. Parameter-free results are obtained
in terms of physical masses and $f=90$ MeV \cite{Granada,KL03}.} \label{fig2}
\end{figure}

\begin{figure}[t]
\begin{center}
\includegraphics[width=12.0cm,clip=true]{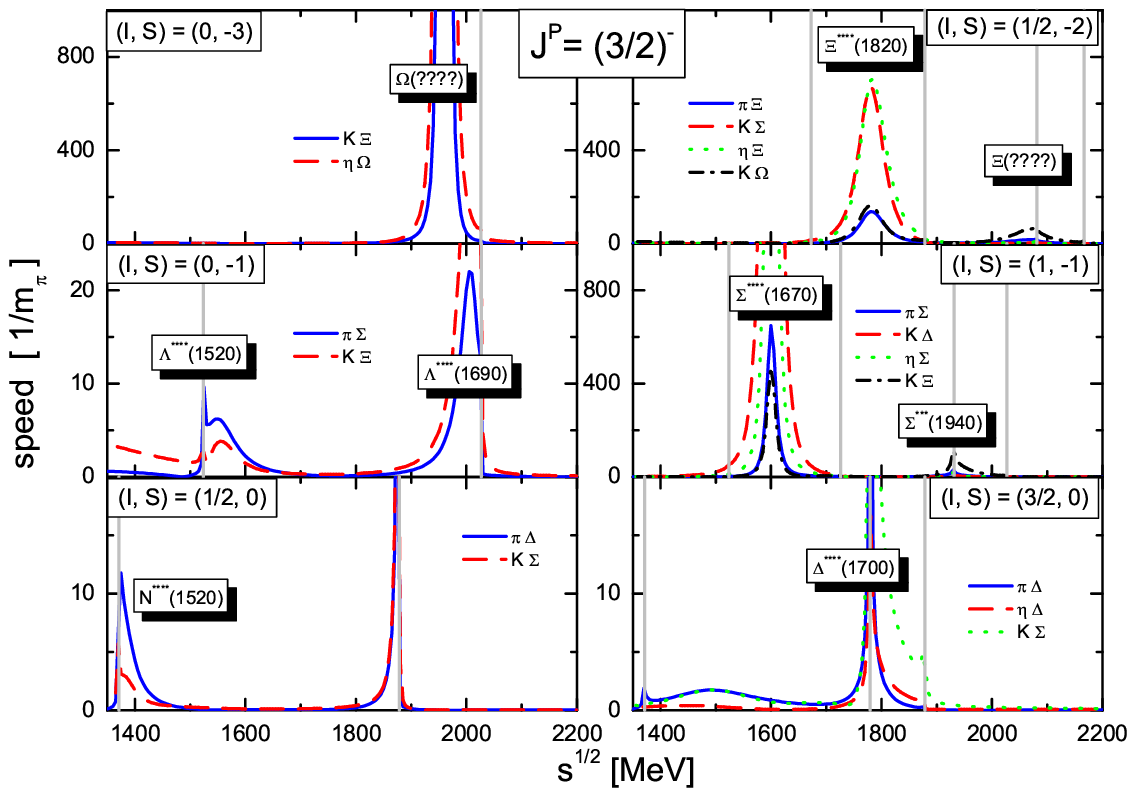}
\end{center}
\caption{Diagonal speed plots of the $J^P=\frac{3}{2}^-$ sector. The vertical lines show the
opening of inelastic meson baryon-decuplet channels. Parameter-free results are obtained
in terms of physical masses and $f=90$ MeV.} \label{fig1}
\end{figure}

Using physical meson and baryon masses the bound-state turn into
resonances as shown in Figs. \ref{fig2},\ref{fig1}. In \cite{KL03} we generalized the notion of a
speed \cite{Hoehler:speed} to the case of coupled-channels in way that the latter reveals the coupling strength
of a given resonance to any channel, closed or open. If a resonance with not too large decay
width sits in the amplitude a clear peak structure emerges in the speed plot even if the
resonance structure is masked by a background phase. In the case of s-wave resonances thresholds
induce square-root singularities which should not be confused with a resonance signal.

The speed plots of Fig. \ref{fig2} show evidence for the formation of the $\Xi(1690)$,
$\Lambda(1405)$, $\Lambda(1670)$ and $N(1535)$ resonances close to their empirical masses. An additional
$(I,S)=(0,-1)$ state, mainly a SU(3) singlet \cite{Jido03,Granada}, can be found as a complex
pole in the scattering amplitude close to the pole implied by the $\Lambda(1405)$ resonance.
There is no clear signal for $(I,S)=(1,-1)$ resonances at this leading order calculation. However,
chiral corrections lead to a clear signal in this sector \cite{LK02} suggesting a state that
may be identified with the  $\Sigma(1750)$ resonance, the only well established s-wave resonance
in this sector. The fact that a second resonance with $(I,S)=(\frac{1}{2},0)$ is not seen in
Fig. \ref{fig2}, even though the 'heavy' SU(3) limit suggests its existence, we take as a
confirmation of the phenomenological observation \cite{LWF02} that the $N(1650)$ resonance
couples strongly to the $\omega_\mu \,N$ channel not considered here.

In Fig. \ref{fig1} speed plots of the $J^P=\frac{3}{2}^-$
sector are shown for all channels in which octet and decuplet resonance states are expected.
It is a remarkable success of the $\chi$-BS(3) approach that it predicts the
four star hyperon resonances $\Xi(1820)$, $\Lambda(1520)$, $\Sigma (1670)$ with masses quite
close to the empirical values. The nucleon and isobar resonances $N(1520)$ and $\Delta (1700)$
also present in Fig. \ref{fig1}, are predicted with less accuracy. The important result here is the
fact that those resonances are generated at all. It should not be expected to obtain already
fully realistic results in this leading order calculation. For instance chiral correction
terms are expected to provide a d-wave $\pi \,\Delta$-component of the $N(1520)$.
We continue with the peak in the (0,-3)-speeds at mass 1950 MeV.
Since this is below all thresholds it is in fact a bound state. Such a state
has so far not been observed but is associated with a decuplet resonance \cite{Schat}.
Further states belonging to the decuplet are seen in the $(\frac{1}{2},-2)$- and
$(1,-1)$-speeds at masses 2100 MeV and 1920 MeV. The latter state can be identified with the
three star $\Xi (1940)$ resonance. Finally we point at the fact that the $(0,-1)$-speeds show
signals of two resonance states consistent with the existence of the four star resonance
$\Lambda(1520)$ and $\Lambda(1690)$ even though in the 'heavy' SU(3) limit we observed only
one bound state. It appears that the SU(3) symmetry breaking pattern generates the 'missing'
state in this particular sector by promoting the weak attraction of the 27-plet contribution
in (\ref{8-10-decom}).

\section{Selfconsistent strangeness propagation in cold nuclear matter}

\begin{figure}[t]
\begin{center}
\includegraphics[width=11.0cm,clip=true]{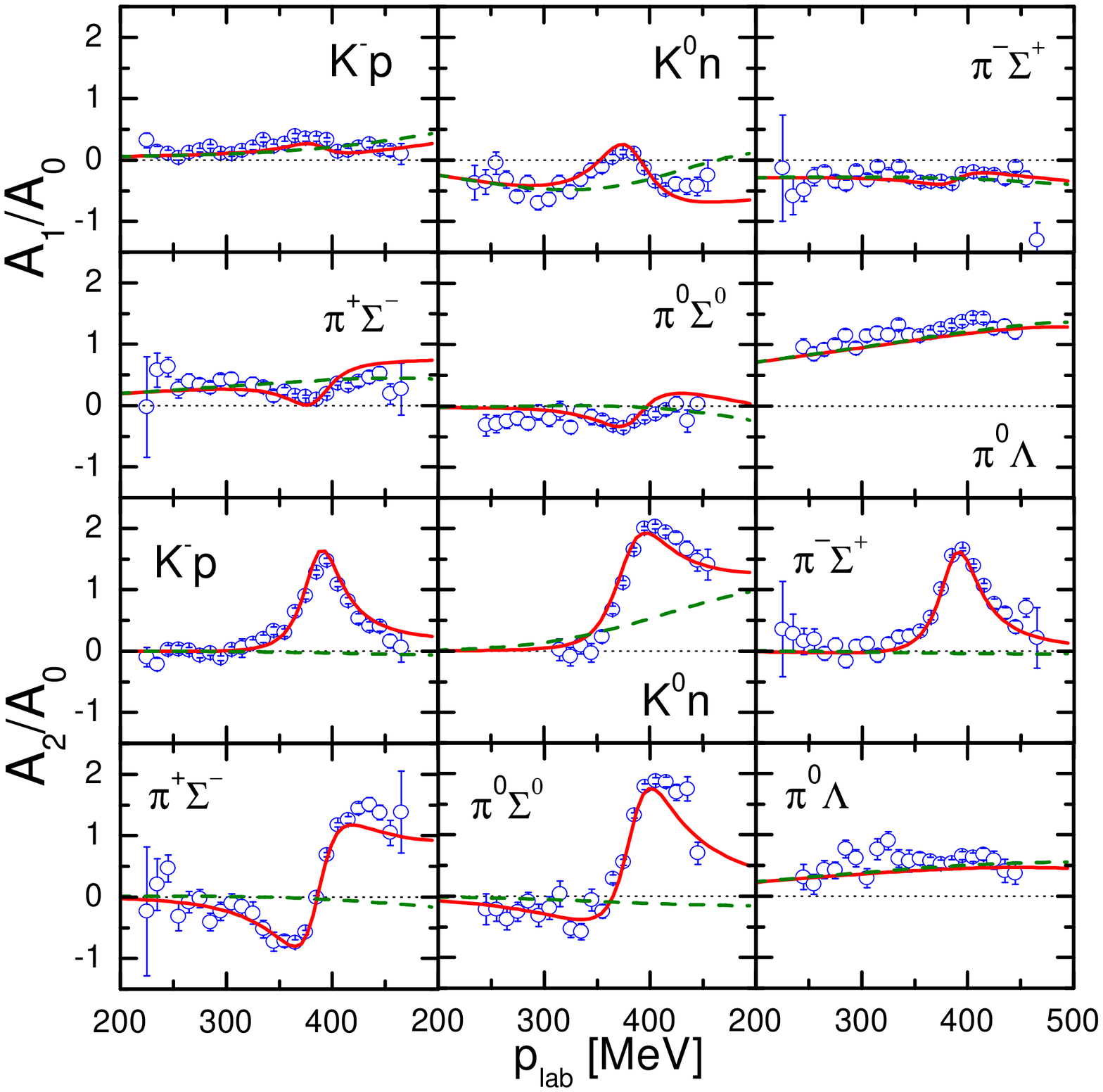}
\end{center}
\caption{Coefficients $A_1$ and $A_2$ for the $K^-p\to \pi^0 \Lambda$,
$K^-p\to \pi^\mp \Sigma^\pm$
and $K^-p\to \pi^0 \Sigma$ differential cross sections. The data are
taken from \cite{mast-pio,bangerter-piS}. The solid lines are the result of the $\chi$-BS(3) approach
with inclusion of the d-wave resonances. The dashed lines show the effect of switching off d-wave contributions.}
\label{fig:a}
\end{figure}

\begin{figure}[t]
\begin{center}
\includegraphics[width=12cm,clip=true]{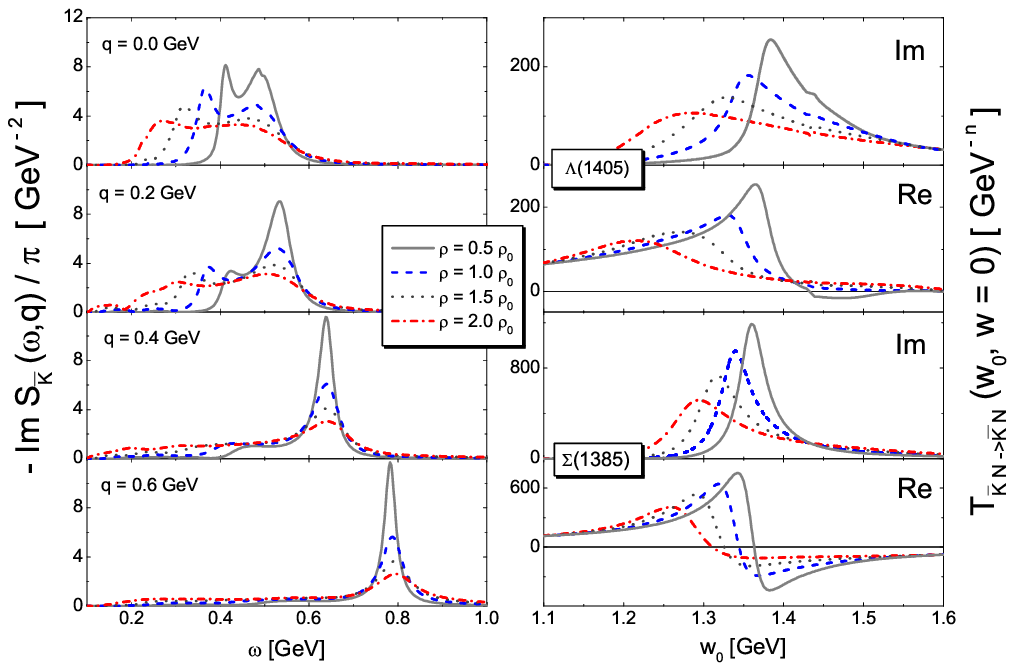}
\end{center}
\caption{The antikaon spectral function is shown in the left-hand panel as
a function of the antikaon energy $\omega$, the momentum $q$ and the
nuclear density with $\rho_0 = 0.17$ fm$^{-3}$. The right-hand panel
illustrates the in-medium modification of the  $\Lambda(1405)$ and
$\Sigma (1385)$ hyperon resonances. It is plotted the real and imaginary
parts of the antikaon-nucleon scattering amplitudes in the appropriate channels.
The hyperon energy and momentum are $w_0$ and $w =0$ respectively.}
\label{fig:kaon-sp}
\end{figure}

We turn to antikaon and hyperon resonance propagation in cold nuclear matter.
The quantitative evaluation of the antikaon spectral function in nuclear matter is a
challenging problem. It should be based on a solid understanding of the antikaon-nucleon
scattering process in free space. Based on a description of the data set one obtains
a set of antikaon-nucleon scattering amplitudes. The present data set for antikaon-nucleon
scattering leaves much room for different theoretical extrapolations to subthreshold energies
\cite{A.D.Martin}. Thus it is of crucial importance to apply effective field theory methods in
order to control the uncertainties. In particular constraints from crossing symmetry and chiral
symmetry should be taken into account. Since the accuracy of the data improves dramatically as the
energy increases it is desirable to incorporate contributions from higher partial waves into
the analysis.
Important information on the p-wave dynamics is provided by angular distributions for the inelastic
$K^-p$ reactions. The available data are represented
in terms of coefficients $A_n$ characterizing the differential cross section
$d\sigma(\cos \theta , \sqrt{s}\,) $ as functions of the center of mass
scattering angle $\theta $ and the total energy $\sqrt{s}$:
\begin{eqnarray}
\frac{d\sigma (\sqrt{s}, \cos \theta )}{d\cos \theta }  &=&
\sum_{n=0}^\infty A_n(\sqrt{s}\,)\,P_n(\cos \theta ) \,.
\label{a-b-def}
\end{eqnarray}
In Fig.~\ref{fig:a} we compare the empirical ratios $A_1/A_0$ and $A_2/A_0$ with the results of
the $\chi$-BS(3) approach. A large $A_1/A_0$ ratio
is found only in the $K^-p\to \pi^0 \Lambda$ channel demonstrating the importance of
p-wave effects in the isospin one channel. The dashed lines of Fig.~\ref{fig:a}, which are obtained when
switching off d-wave contributions, illustrate the importance of the $\Lambda(1520)$
resonance for the angular distributions in the isospin zero channel. Note also the sizeable
p-wave contributions at somewhat larger momenta seen in the charge-exchange reaction
of Fig.~\ref{fig:a}.

In Fig. \ref{fig:kaon-sp} we present the antikaon spectral function together with
the antikaon-nucleon scattering amplitudes of selected channels at various nuclear
matter densities. The results are based on antikaon-nucleon scattering
amplitudes obtained within the chiral coupled-channel effective field theory
\cite{LK02}, where s-, p-and d-wave contributions were considered. The
many-body computation \cite{LuKor} was performed in a self consistent manner respecting
in addition constraints arising from covariance. The antikaon spectral function exhibits
a rich structure with a pronounced dependence on the antikaon three-momentum. That reflects
the coupling of the $\Lambda (1405)$ and $\Sigma (1385)$ hyperon states to the $\bar K N$
channel. Typically the peaks seen are quite broad and not always of quasi-particle type.
The figure illustrates that at zero momentum the spectral
function acquires a rather broad distribution as the nuclear density increases, with support
significantly below the free-space kaon mass. At moderate momenta $q= 400$ MeV the spectral
function looks significantly different as compared to the one at $q= 0$ MeV. It is characterized
by a peak at close to $\omega = \sqrt{m_K^2+q^2}$ and a pronounced low-energy tail. As the
density increases the strength in the peak diminishes shifting more and more strength into the low-energy tail.
As was emphasized in \cite{ml-sp,LuKor} the realistic evaluation of the antikaon propagation in nuclear
matter requires the simultaneous consideration of the hyperon resonance propagation.
The most important contributions, the s-wave $\Lambda (1405)$ and p-wave $\Sigma (1385)$
resonances, experience important medium modifications as demonstrated in Fig.
\ref{fig:kaon-sp}. The results at $2 \,\rho_0$ should be considered cautiously because
nuclear binding and correlation effects were not yet included in \cite{LuKor}.

\section{Summary}

In this talk we reported on recent progress in the understanding of baryon resonances
based on chiral-coupled channel dynamics. The reader was introduced to
an effective field theory formulation of chiral coupled-channel dynamics. Leading order
results predict the existence of s- and d-wave baryon resonances with a spectrum remarkably
close to the empirical pattern without any adjustable parameters. The formation of
resonances is a consequence of the chiral SU(3) symmetry of QCD, i.e. in an effective field
theory, that was based on the chiral SU(2) symmetry only, no resonances would be formed.

As a further application of chiral coupled-channel dynamics results for antikaon and hyperon
resonance propagation in cold nuclear matter were presented. Realistic scattering amplitudes
that are consistent with empirical differential cross sections were obtained after
including chiral corrections terms systematically. The spectral function of the antikaon
shows a strong momentum and density dependence. For the $\Lambda(1405)$ and $\Sigma(1385)$
resonances attractive mass shifts are predicted.

\end{document}